\documentclass[conference]{IEEEtran}
\IEEEoverridecommandlockouts

\usepackage{cite}
\usepackage{amsmath,amssymb,amsfonts}
\usepackage{algorithmic}
\usepackage{graphicx}
\usepackage{textcomp}
\usepackage{xcolor}
\def\BibTeX{{\rm B\kern-.05em{\sc i\kern-.025em b}\kern-.08em
    T\kern-.1667em\lower.7ex\hbox{E}\kern-.125emX}}

\usepackage{fancyhdr}

\usepackage{balance}

\newcommand{\equals}{\!=\!}
\newcommand{\lteq}{\!\le\!}
\newcommand{\minus}{\!-\!}

\newcommand{\gthan}{\!>\!}
\newcommand{\lthan}{\!<\!}

\newcommand{\noteq}{\! \ne \!}

\begin{document}

\title{Leveraging Wake-Up Radios in UAV-Aided LoRa Networks: Some Preliminary Results\\ on a Random-Access Scheme\\
\thanks{This work was supported by the Science
and Engineering Research Board (SERB), Government of India, through its
Start-up Research Grant (SRG) under SRG/2020/001491.}
}

\author{\IEEEauthorblockN{Shashank M. S. Panga}
\IEEEauthorblockA{\textit{School of Electrical Sciences} \\
\textit{Indian Institute of Technology Bhubaneswar}\\
Khordha 752050, India \\
pss14@iitbbs.ac.in}
\and
\IEEEauthorblockN{Siddhartha S. Borkotoky}
\IEEEauthorblockA{\textit{School of Electrical Sciences} \\
\textit{Indian Institute of Technology Bhubaneswar}\\
Khordha 752050, India \\
borkotoky@iitbbs.ac.in}
}

\maketitle

\thispagestyle{fancy}
\begin{abstract}
We present a transmission scheme aimed at integrating wake-up radios (WuR) into LoRa sensor networks featuring UAV-mounted gateways. The sensors are informed of the UAV's arrival with the help of WuR, followed by sensor-to-UAV data transfer with frequency and spreading-factor hopping. The proposed scheme provides significant energy savings at the sensors while maintaining similar reliability levels compared to a scheme in which Class B LoRa beacons are used to perfectly synchronize the sensors with the UAV's data-collection window.  
\end{abstract}


\section{Introduction}
\label{sec:Intro}
In the recent years, Long Range (LoRa) networks have been extensively employed in sensor networks where the sensors transmit small amounts of data (e.g., a few bytes) relatively infrequently (e.g., every few minutes or hours) over long distances (up to several kilometers)~\cite{MTP23}. As with any wireless communication system, there is a trade-off between coverage and power consumption in LoRa. Long distance transmissions must  be made with higher power, or higher spreading factors (thus longer frames and higher energy expenditure), or both~\cite{GeR17}. Consequently, it can be of tremendous benefit to employ a UAV-mounted gateway to periodically hover over the sensors, which can transmit their data over a relatively short distance to the UAV with lower energy consumption. The UAV can deliver the sensor data over a high-capacity link (e.g., LTE or 5G) to the final destination, which may be a control station that processes the data for analysis, decision making, and control.  This possibility has motivated several prior research works that employ UAVs in LoRa networks~\mbox{\cite{XLZ23}--\!\cite{CHW18}}.

Along similar lines, this paper considers UAV-aided data transfer  from a cluster of LoRa-equipped sensor nodes or end devices (EDs) to a control station. Typical examples of target applications include  environmental monitoring use cases, where the EDs may measure parameters such as air quality, ambient temperature, soil moisture, etc from time to time. 
While the use of an UAV reduces the  energy consumption at the EDs, it imposes the additional requirement of informing the EDs about the arrival and hovering times of the UAV, so that they transmit only when the UAV is in the vicinity. However, this aspect has been largely ignored in prior literature, under the implicit assumption that the EDs have the necessary knowledge.

One practical way to impart that knowledge to the EDs is to send control messages from the control station. This can be accomplished using the Class B functionality of LoRaWAN~\cite{ClassB}, which allows the transmission of periodic beacon messages for synchronization and ping messages for transferring the control information. 
While effective, this approach increases the energy consumption at the EDs, which must scan for and demodulate the beacons and pings at regular intervals.  An alternative possibility  involves the use of wake-up radios (WuR)~\cite{PMK17}. A WuR module, which can be installed in an ED in addition to its main radio module, is characterized by its extremely low power consumption while listening for and receiving signals. This feature enables an ED to completely power off its main radio during idle times while letting the WuR module scan for incoming wake-up beacons (WUB). An WuR enabled gateway that wishes to establish communication with the ED can send a WUB, upon whose receipt the ED powers on its main radio to send or receive data. This is inherently attractive for UAV communications, since the WUBs can be used for alerting the EDs when the UAV arrives, and send small amounts of information about the UAV's data collection window. The integration of LoRa and WuR for non-UAV settings has been explored in prior works~\mbox{\cite{AGM18}--\!\cite{PMM18}}. These works involve the selection of an ED node as the cluster head tasked with waking up the other EDs using WuR upon instruction from the gateway.

The objective of this work is to investigate the use of WuRs in a UAV-aided setting for LoRa-based delay-tolerant applications,  and to design a low-complexity uplink transmission scheme for this purpose. Unlike~\cite{AGM18}--\!\cite{PMM18}, our approach  avoids a cluster head; instead, the WuR-equipped UAV transmits WUBs to wake up the EDs. In our design and analysis, we account for the practical constraint that a WUB is not guaranteed to reach all EDs; instead, the wake-up event for each ED has a  probabilistic nature. The wake-up probability  depends, for example, on the time-varying channel conditions on the WuR link from the UAV to the ED, and on the length of the link. Thus, given a sequence of WUB transmissions by the UAV, different EDs may wake up at different times. Since the UAV  hovers over the EDs only for a limited duration, the EDs that wake up late may not have enough time to send all their data to the UAV. Our main contributions are:
\begin{enumerate}
    \item We propose a mechanism for WuR- and UAV-aided data collection in LoRa  networks, in which an ED powers off its LoRa module until awaken by a WUB from the UAV. The ED then transmits its data to the UAV over a time slotted channel, using a random frequency and spreading factor for each transmission. Messages not transmitted during the UAV's data collection window due to the UAV's departure are sent directly to the control station exploiting LoRa's long-range capabilities.  

    \item We analytically derive the message delivery performance and energy expenditure under   the scheme.

    \item We provide  comparisons of the WuR-aided scheme with a benchmark mechanism that employs class B LoRa beacons for synchronization, and with another scheme that does not employ UAV and WuR.
\end{enumerate}

\pagestyle{empty}

\section{System Model}
\label{sec:sys_mod}



We consider a cluster of $n$ LoRa EDs. An UAV positions itself above the cluster for collecting data from the EDs, and transporting them via a high-capacity link to a control station. The data are represented as $b$-byte messages, each message corresponding to a sensor measurement. The number of messages an ED has accumulated for transmission at the time of the UAV's arrival is a discrete uniform random variable that takes values between 1 and $M_{\max}$. This may correspond to a network in which equal fractions of EDs perform measurements with periodicity \mbox{$T_u$, $2T_u$, $\ldots$, $M_{\max}T_u$,} where $T_u$ is the time between two successive arrivals of the UAV. 

The EDs and UAV are equipped with LoRa and WuR modules. The WuR is used for waking up the EDs upon the UAV's arrival, whereas the LoRa modules are used by the EDs to send data to the UAV, and if needed, directly to the control station. The probability that a WUB from the UAV successfully wakes up an ED is denoted by $P_b$. For each uplink transmission, the ED chooses one out of $N_f$ non-overlapping frequency bands uniformly at random.  The transmit power from ED to UAV is denoted by $\mathcal{P}_t$, and from an ED to the control station by $\tilde{\mathcal{P}}_t$.

\subsection{LoRa Transmission Characteristics}
LoRa signals employ the chirp spread spectrum (CSS) modulation. A CSS signal is characterized by its spreading factor (SF), which is an integer between $7$ and $12$. Higher SFs provide greater immunity against thermal noise at the expense of a longer frame duration. We denote the receiver sensitivity while using SF $k$ by $\zeta(k)$. The duration of a frame that employs SF $k$, a channel code of rate $4/5$, and  no optional header field can be expressed as
\begin{align}  
    L_f(k) &= \bigg[20.25 +  
         \max\left\{ 5\left\lceil \frac{2b - k + 11}{k \minus 2q} \right\rceil, 0\right\}\bigg]\,\frac{2^{k}}{w}, 
\end{align}
where $w$ is the transmission bandwidth and $q$ is 1 when low data rate optimization is enabled and 0 otherwise~\cite{Sem13}. 

The spreading factors are quasi-orthogonal in the sense that a frame using SF $k$ is highly likely to survive a collision with a frame that employs SF $k' \noteq k$ unless the latter has a much higher power that the former. For example, as per data in~\cite{MSG19}, the lost frame in an inter-SF collision event must be weaker than the colliding frame by $-8$~dB to $-25$~dB depending on the SF.  For simplicity, we assume here that multiple frames arriving simultaneously over the same frequency channel with different spreading factors are all correctly received, whereas if multiple frames with the same spreading factor arrive simultaneously, then all are lost. This is a reasonable model under the assumption that the EDs are not very far apart such that there distances to the UAV are comparable (resulting in similar path loss) and the impact of multipath fading on the ED-to-UAV links are mild owing to a strong line-of-sight component (so that the received power in two frames differing by a margin as large as $-8$~dB or higher is negligible.)

Let $K_d$ denote the SF used for direct transmissions to the control station, which is assumed to be far enough from the EDs so that $K_d$ is one of the larger SFs (e.g., $11$ or $12$). The success probability on the direct link is denoted by $P_d$, which takes into account the impact of small scale fading, large scale path loss, as well as interference.

\section{Data Collection Scheme}
\label{sec:scheme}
We consider a slotted communication system with slot length $T_s$, and suppose the UAV hovers over the EDs for a duration of $N_s$ slots, numbered 0 through $N_s \minus 1$. Slot 0 begins once the UAV positions itself above the cluster. At the beginning of each slot, the UAV broadcasts a WUB that  contains a short data field containing the current slot number $i$ and the value of $N_s$. Now consider an ED that wakes up in slot $i$, and define 
\begin{align}
    N(i) \equals N_s \minus i.
\end{align}
Thus the ED has $N(i)$ slots to send its data. Suppose it has $M$ messages to send. If $M \leq N(i)$, then the ED selects $M$ out of $N(i)$ slots at random and transmits its messages during these slots. If $M \gthan N(i)$, then the ED randomly selects $N$ messages and send them over the $N(i)$ slots. The remaining $M \minus N(i)$ messages are transmitted directly to the control station.

For every transmission to the UAV, the SF is chosen uniformly at random, from the set $\mathbf{K} \equals \{7,8,\ldots,K_m\}$, where $7 \lteq K_m \lteq 12$. Thus, the probability of selecting SF $k \in \mathbf{K}$ is 
\begin{align}
    \eta_k = \frac{1}{K_m-6}.
\end{align}
Each slot is just long enough to accommodate the longest frame, which corresponds to SF $K_m$. Thus, 
\begin{align}
    T_s = L_f(K_m).
\end{align}

For direct transmission to the control station with SF $K_d$, the ED employs a backoff mechanism by transmitting the frame $Z$ slots after the UAV has left, where $Z$ is a random integer uniformly distributed over $\{0,1,\ldots,z_m\}$. We assume $z_m$ to be sufficiently large so as to keep the probability of collisions among directly transmitted messages small. This allows us to model the direct link as one with a constant erasure probability $1 \minus P_d$ (resulting from fading, shadowing, and ambient interference from other networks) rather than as a channel whose characteristics vary dynamically based on the direct transmission behavior by the EDs.      

\subsection{Chirp Spread Spectrum  vs. On-Off Keying}
In addition to CSS, most commercial LoRa devices include an on-off keying (OOK) module, which is suitable for short-range communications. Our approach, however, employs the CSS module even on the short-distance ED-to-UAV uplink. Unlike OOK, the CSS frames provide six near-orthogonal SFs. Since the EDs transmit without prior resource allocation within a short time span, using CSS enables them to hop over the SFs to reduce the collision-induced losses. Exploration of alternatives employing OOK is left as future work.

\begin{table}
    \centering
    \caption{Important symbols and abbreviations}
    \begin{tabular}{c|c}
        Symbol & Definition \\ \hline
        SF & Spreading factor\\
        WuR & Wake-up radio\\
        WUB & Wake-up beacon \\
        $n$ & No. of EDs in the cluster \\
        $M_{\max}$ & Max. possible measurements at an ED\\
        $P_M(m_0)$ & Prob. that an ED has $m_0$ messages to send\\
        $P_W(i)$ & Prob. that an ED wakes up in slot $i$\\
        $N_f$ & Number of frequency bands \\
        $\mathcal{P}_t$ & Transmit power on ED-to-UAV uplink \\
        $\tilde{\mathcal{P}_t}$ & Transmit power for direct tx. to controller \\
        $L_f(k)$ & LoRa Frame length for SF $k$ \\
        $N_s$ & No. of slots over which the UAV receives\\
        $T_s$ & Slot length\\
        $P_b$ & Probability a WUB is successfully received\\
        $\mathbf{K}$ & Set of SFs used for transmissions to the UAV\\
        $K_m$ & Largest SF used on ED-to-UAV link\\
        $\eta_k$ & Prob. that SF $k$ is used for a frame\\
        $K_d$ & SF for direct tx. from ED to controller \\
        $\lambda$ & Probability a message is not sent to UAV\\
        $\mathcal{T}(s)$ & Probability a message is sent in slot $s$\\
        $T_u$ & UAV arrival period \\
        $T_p$ & Ping message periodicity for Class B\\
        $T_b$ & Beacon message periodicity for Class B\\
        $L_p$ & Ping message airtime for Class B\\
        $L_b$ & Beacon message airtime for Class B\\
    \end{tabular}
    \label{tab:my_label}
\end{table}

\section{Performance Analysis}
\label{sec:analysis}
We now derive the \textit{message delivery probability} (MDP), which is the  probability that an arbitrary message from an ED is delivered to the control station, either via the UAV or over the direct link. The MDP can  be expressed as
\begin{align}
    \mathcal{S} = \mathcal{S}^{\mathrm{(UAV)}} + \mathcal{S}^{\mathrm{(Direct)}},
\end{align}
where $\mathcal{S}^{\mathrm{(UAV)}}$ is the probability that the message was delivered to the UAV gateway and $\mathcal{S}^{\mathrm{(Direct)}}$ is the probability that the message was not transmitted to the UAV due to shortage of slots (either because $N_s$ is small, or because the ED woke up late into the data-collection window), but the direct transmission to the control station is successful. 

We begin the derivation by noting that the probability mass function for the slot in which an ED wakes up (denoted by $W$) is 
\begin{align}
    P_W(i)  =
    \begin{cases}
        (1-P_b)^{i} P_b, \quad  &0 \leq i \leq N_s-1 \\
        0,\quad &\text{otherwise}.
    \end{cases}
\end{align}
Observe that if an ED with $m_0$ messages wakes up in slot $i$, then the probability that it transmits a frame to the UAV in slot $s$ is
\begin{align} \label{eq:cond_tx_prob}
    P_{\mathrm{tx}}(s|W=i,M=m_0) =  
    \begin{cases}
        \min\left\{\frac{m_0}{N(i)}, 1\right\}, \quad &s \geq i \\
        0 \quad &s < i.
    \end{cases}
\end{align} 

Now we proceed to characterize the collision probability on the ED-to-UAV uplink.  Suppose our desired frame (i.e., the frame carrying the message whose success probability is to be determined) is sent in slot $s$. The probability that it collides with a frame from a given -- but arbitrary -- ED is (see Appendix~\ref{app:coll_prob} for the derivation)
\begin{align} \label{eq:coll_prob}
P_{\mathrm{col}}(s) 
&= \frac{1}{M_{\max}}  \sum_{m_0 = 1}^{M_{\max}} \sum_{i=0}^{s}\min\left\{\frac{m_0}{N(i)}, 1\right\} P_W(i).
\end{align}
To enforce a loss, the collision must occur over the same channel and the colliding frame must use the same SF as the desired frame. Hence, we obtain the probability of frame loss in slot $s$ due to one specific ED as
\begin{align}
    L^{(1)}(s) = \frac{\eta_k}{N_f}P_{\mathrm{col}}(s).  
\end{align}
Since there are a total of $n$ EDs in the cluster, we can express the probability of successful reception of a frame sent to the UAV gateway in slot $s$ to be  
\begin{align} \nonumber
    P_{\mathrm{succ}}^{\mathrm{(UAV)}}(s) &= (1- L^{(1)}(s))^{n-1} \\
    &= \left(1- \frac{ \eta_k}{N_f}P_{\mathrm{col}}(s)\right)^{n-1}. 
\end{align}
The probability that the desired message is sent over slot $s$ is given by (see Appendix~\ref{app:des_tx_prob} for derivation)
\begin{align} \label{eq:des_tx_prob}
    \mathcal{T}(s) = \sum_{m_0=1}^{M_{\max}} \sum_{i=0}^{s} \frac{P_W(i)}{M_{\max}}  \min\left\{\frac{N(i)}{m_0},1\right\} \frac{1}{N(i)}.
\end{align} 
The success probability on the UAV uplink is thus
\begin{align}
    \mathcal{S}^{\mathrm{(UAV)}} &= \sum_{s=0}^{N_s-1} \mathcal{T}(s) P_{\mathrm{succ}}^{\mathrm{(UAV)}}(s).
\end{align}

For the direct transmission to the control station,
\begin{align} 
    \mathcal{S}^{\mathrm{(Direct)}} &= \lambda P_d,
\end{align}
where $\lambda$ is the probability that the message was not picked for transmission to the UAV gateway, thus leading to its direct transmission to the control station. It is given by (see Appendix~\ref{app:lambda} for derivation)
\begin{align} \label{eq:lambda}
    \lambda 
    &=  1 - \sum_{m_0=1}^{M_{\max}} \sum_{i=0}^{N_s-1} \frac{P_W(i)}{M_{\max}}  \min\left\{\frac{N(i)}{m_0},1\right\}.
\end{align}

The average energy consumption at an ED per message is easily shown to be
\begin{align}
    \mathcal{E} = (1-\lambda)\mathcal{P}_t \sum_{k\in \mathbf{K}} \eta_k  L_f(k) + \lambda \tilde{\mathcal{P}}_t L_f(K_d).
\end{align}

\vspace{2mm}
\section{Baseline Schemes}
\label{sec:baselines}
For comparison, we consider two baseline schemes:
\subsection{Direct Transmission to Control Station}
This scheme does not employ an UAV; instead, the EDs transmit directly to the control station using SF $K_d$ (the same SF as used by the WuR-aided scheme for direct transmissions). The message delivery probability in this case is simply the success probability on the direct link, that is,
\begin{align}
    \mathcal{S}^{\mathrm{(No UAV)}} = P_d.
\end{align}
The energy consumption per message at an ED is 
\begin{align}
    \mathcal{E}^{\mathrm{(No UAV)}} = \tilde{\mathcal{P}}_t L_f(K_d).
\end{align}


\subsection{UAV-Aided Data Collection With Ideal Class-B LoRa Synchronization}
This scheme employs an UAV for data collection, but no WuR is used. Instead, the EDs operate as Class B LoRa devices. The LoRa gateway at the control station transmit periodic beacons to both the UAV and the EDs so that all of them are time-synchronized. The ping slots are used to inform the EDs of the time the UAV is expected to arrive. Because this is a benchmark protocol, we assume there is no loss of beacon and ping messages (ideal Class B operation) and hence all EDs are aware of the UAV's arrival time. As in the UAV-based scheme, the EDs randomly choose slots to transmit their messages, transmitting directly to the control station the messages that could not be accommodated in a slot. The analysis of ED is a special case of the analysis in Section~\ref{sec:analysis}, with $P_b \equals  P_W(0) \equals 1$, and $N(0) = N_s$. 
As a result, the message delivery probability becomes
\begin{align}
    \mathcal{S}^{\mathrm{(Class B)}} = (1-\lambda') \left(1- \frac{ \eta_k}{N_f}P_{\mathrm{col}}' \right)^{n-1} + \lambda' P_d,
\end{align}
where $P_{\mathrm{col}}'$ and $\lambda'$ are given by
\begin{align}
    P_{\mathrm{col}}' = \frac{1}{M_{\max}}  \sum_{m_0 = 1}^{M_{\max}} \min\left\{\frac{m_0}{N_s}, 1\right\}
\end{align}
and
\begin{align}
    \lambda' = 1 - \frac{1}{M_{\max}} \sum_{m_0=1}^{M_{\max}}   \min\left\{\frac{N_s}{m_0},1\right\},
\end{align}
which are found using similar reasoning as employed in Section~\ref{sec:analysis} and Appendices~\ref{app:coll_prob}--\ref{app:lambda}.

The average transmission energy per message for an ED is
\begin{align}
    \mathcal{E}^{\mathrm{(Class B)}}_{\mathrm{TX}} &= (1-\lambda')\sum_{k\in \mathbf{K}} \eta_k \mathcal{P}_t L_f(k) + \lambda' \tilde{\mathcal{P}}_t L_f(K_d).
\end{align}
The average energy spent by an ED per cycle (i.e, from one UAV arrival event to the next) in receiving ping and beacon frames is 
\begin{align}  
    \mathcal{E}^{\mathrm{(Class B)}}_{\mathrm{RX}} &= \left(\frac{T_u}{T_p}L_p + \frac{T_u}{T_b}L_b\right)\mathcal{P}_r,
\end{align}
where $T_u$ is the time between successive arrivals of the UAV at the cluster, $T_p$ is the ping period for Class B LoRa beacons, $T_b$ is the beacon period, $L_p$ is the duration of a ping frame, $L_b$ is the duration of a beacon frame, and $\mathcal{P}_r$ is the power consumption associated with receiving downlink signals. 

\section{Numerical Results}
\label{eq:results}

We now provide numerical results on the WuR-aided mechanism vis-{\`{a}}-vis the benchmark protocols. Unless stated otherwise, the simulation parameters are as listed in Table~\ref{tab:sim_values}. All our simulations resulted in numerical values virtually identical to those obtained 
from the analytical models in the preceding sections, hence the graphs that follow are based on the analysis.

\begin{table}
    \centering
    \caption{Default parameter values}
    \begin{tabular}{c|c}
         Parameter & Value \\ \hline
         Number of EDs in the cluster &  30\\
         Possible number of messages at EDs & \{1,2,3,4,5\} \\
         Number of frequency bands & 8\\
         Bytes per message & 10\\
         SF set for ED-to-UAV frames & \{7,8,9,10\} \\
         Tx. power for ED-to-UAV frames & 6~dBm \\
         Number of slots for ED-to-UAV frames & 25\\
         SF on direct link & 11\\
         Tx. power on direct link & 14~dBm\\
         Success prob. on direct link & 0.75\\
         WUB success probability & 0.75\\
         UAV arrival periodicity & 1~hr\\
         Ping periodicity for Class B LoRa & 64~s\\
         Beacon periodicity for Class B LoRa & 128~s\\ 
         Class B ping message size & 4 bytes \\
         Class B beacon message size & 16 bytes
    \end{tabular}
    \label{tab:sim_values}
\end{table}

\begin{figure}
    \centering
    \includegraphics[scale=0.35]{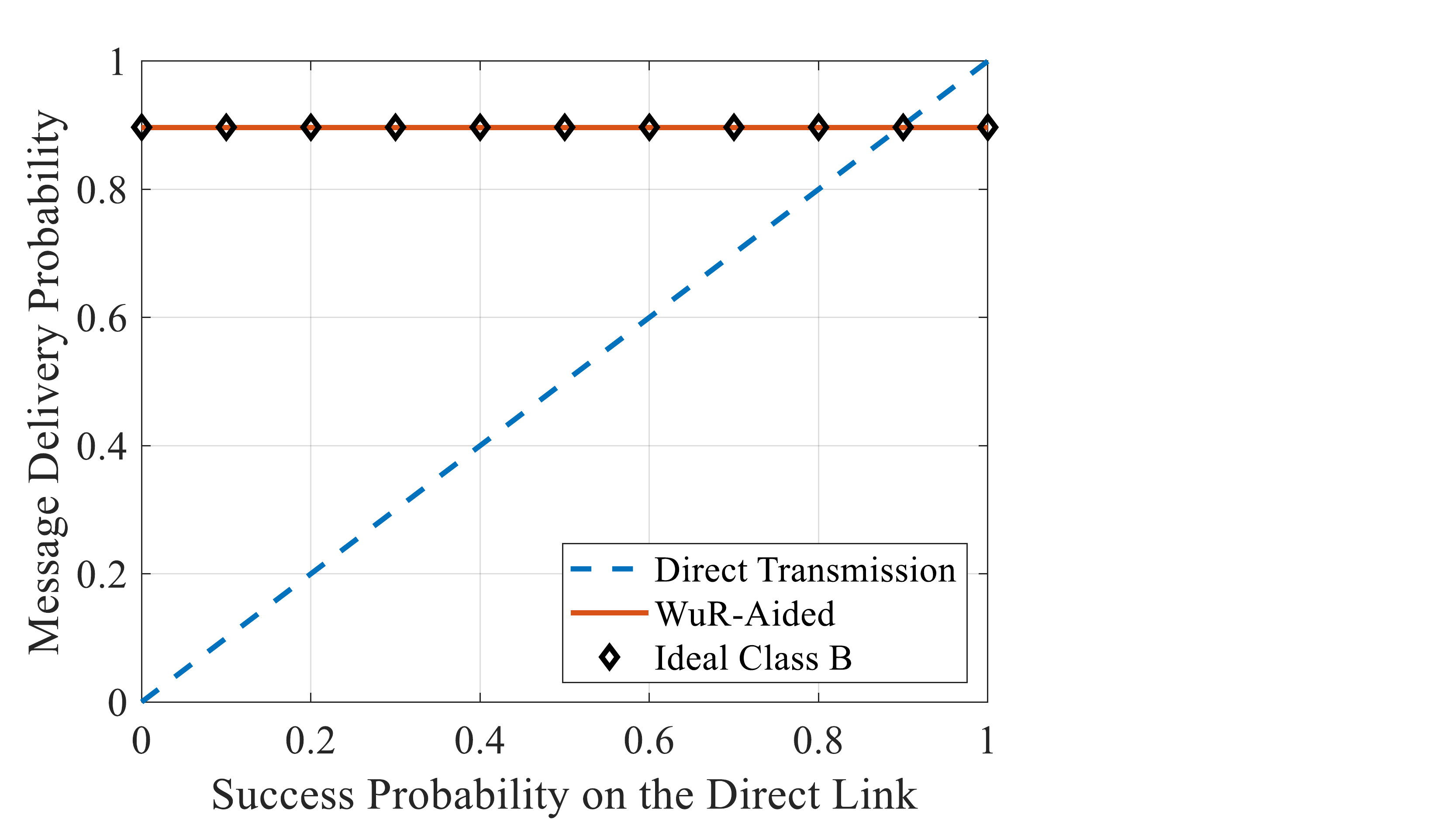}
    \caption{Delivery performance as a function of the direct link's quality.}
    \label{fig:Fig1_MDP_vs_Pd}
\end{figure}

Fig.~\ref{fig:Fig1_MDP_vs_Pd} compares the MDP of the three schemes as a function of the success probability on the direct link from the ED to the control station. The MDP for the direct transmission scheme is simply the success probability on the direct link, as only that link is used for all frames. By contrast, the UAV-based schemes (WuR-aided and ideal Class B) use that link only when a message cannot be accommodated in a slot for transmission to the UAV. As expected,  UAV-based schemes generally perform much better than direct transmission. The only exception is when the direct link is of very high quality; however this is quite unlikely unless the control station is nearby (in which case, one would likely not opt for UAVs anyway); in addition, the good performance would require the use of a large SF, leading to much higher energy consumption at the EDs than the UAV-aided schemes. The key observation from Fig.~\ref{fig:Fig1_MDP_vs_Pd} is that even though the WuR-aided scheme relies on an imperfect wake-up mechanism in which the WUB is not guaranteed to wake up all EDs right after the UAV's arrival, it performs almost identically to the ideal Class B scheme in which we assume beacons from the control station to perfectly synchronize and wake up all EDs right when the UAV arrives, so that all EDs can make use of all $N_s$ slots. Note that a practical Class B scheme would perform worse than our ideal baseline, because not all ping frames may be received, leading to some EDs being unaware of the UAV's arrival time. If the Class B control messages are sent from the control station, then this problem will exacerbate as the downlink quality  from the control station to the EDs deteriorates. In the extreme case where the EDs are not reachable directly from the control station even with the largest SF and maximum allowed downlink transmit power, the Class B scheme will become dysfunctional. The WuR-aided scheme does not suffer from such limitations.           

\begin{figure}
    \centering
    \includegraphics[scale=0.35]{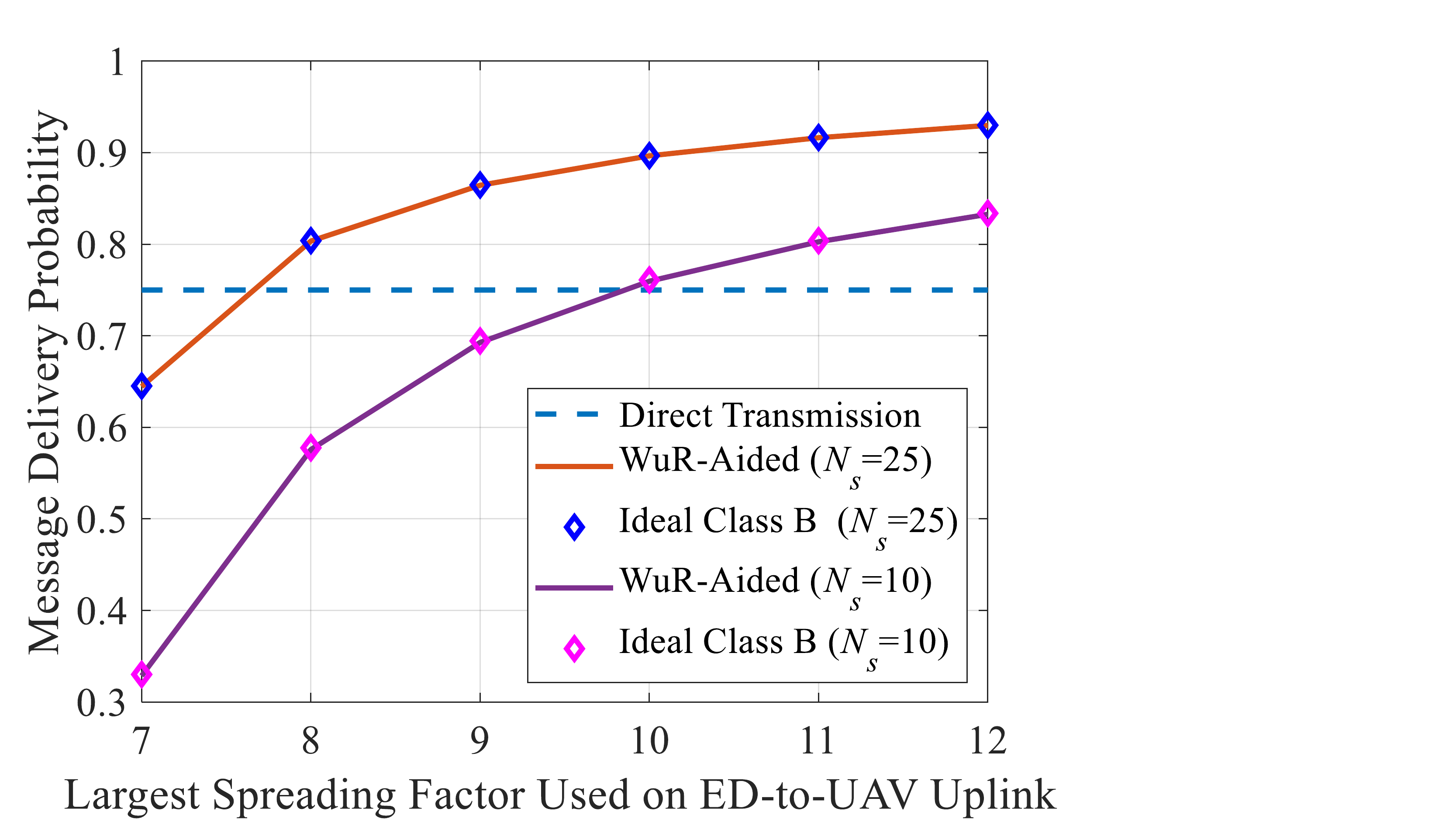}
    \caption{Delivery performance as a function of the SF set.}
    \label{fig:Fig2_MDP_vs_maxSF}
\end{figure}

\begin{figure}
    \centering
    \includegraphics[scale=0.35]{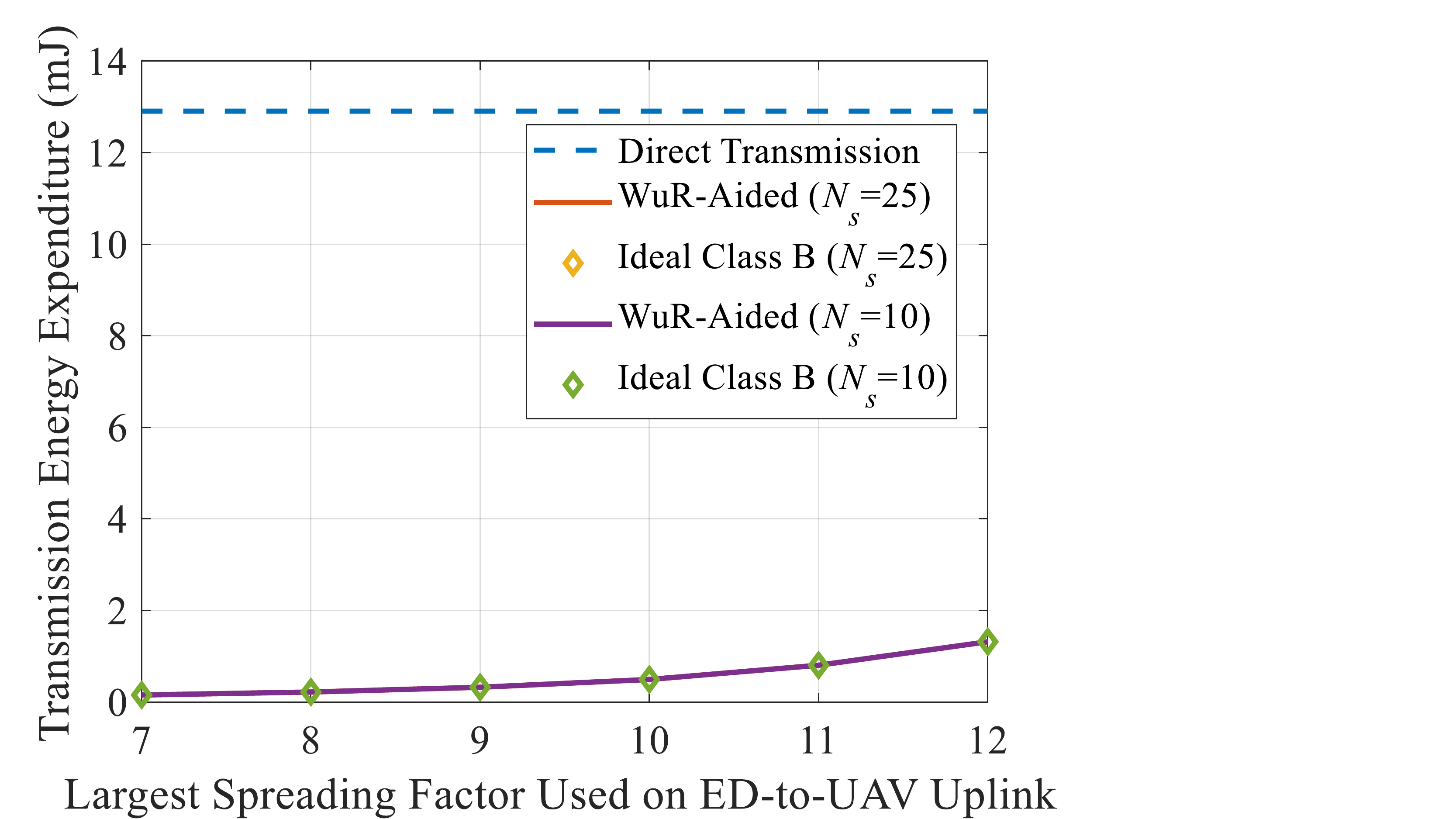}
    \caption{Transmission energy expenditure as a function of the SF set.}
    \label{fig:Fig3_energy_vs_maxSF}
\end{figure}

The impact of the set of SFs used on the ED-to-UAV uplink on the message delivery rate is examined in Fig.~\ref{fig:Fig2_MDP_vs_maxSF}, where the possibilities are varied from the singleton set $\{7\}$ to the set of all six SFs, that is $\{7,8,\ldots,12\}$. With larger number of SFs, the delivery probability improves due to greater number of (quasi-)orthogonal resources. The figure also illustrates the impact of the number of slots allocated for UAV receptions. A larger number of slots result in fewer collisions as the probability of picking a certain slot for a frame decreases, and thus fewer collisions occur. In all instances, we observe that the WuR-aided mechanism performs almost identically to the ideal Class-B scheme.  The corresponding energy expenditures are examined in Fig.~\ref{fig:Fig3_energy_vs_maxSF}, which plots the average transmission energy spent by an ED per message. The transmission energy for the direct transmission scheme is expectedly much higher than the UAV-aided schemes due to the use of SF 11 for all transmissions as opposed to the use of SFs $7$, $8$, $9$, and $10$ in equal proportions in the UAV-aided schemes. The transmission energy incurred in both WuR-aided and the ideal Class B schemes are virtually identical. Note that the transmission energy for the WuR-aided scheme would be exactly the same as for ideal Class B if the WUB reaches each ED early enough to ensure that the UAV's data-collection window has at least $M_{\max}\equals 5$ slots left. If this is condition is not met, then some of the EDs will have to transmit directly to the control station using SF $K_d \equals 11$, leading to longer frames and higher transmission energy expenditure than the ideal Class B scheme. The closeness of the two curves proves that such direct transmissions occur quite rarely to enforce any noticeable increase in the energy expenditure for the WuR-aided scheme. We also observe that both choices of $N_s$ lead to approximately the same transmission energy consumption for the two UAV-aided schemes. This is due to the fact that direct transmissions occur rarely (or never in the case of ideal Class B, unless $N_s \lthan M_{\max}$). 

\begin{figure}
    \centering
    \includegraphics[scale=0.35]{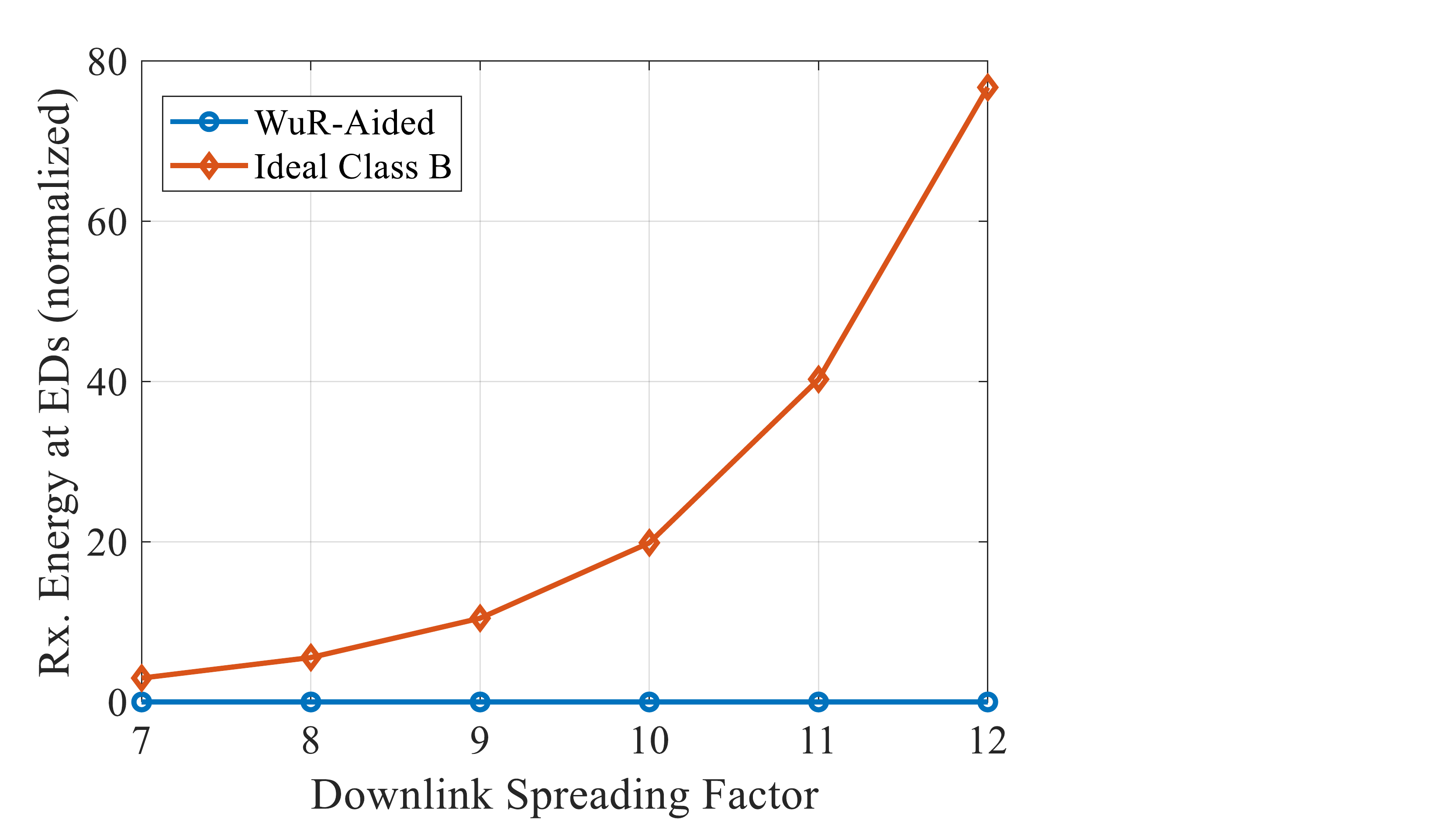}
    \caption{Energy spent in receiving downlink signals.}
    \label{fig:Fig6_RxEnergy}
\end{figure}

In practice, the energy consumption of the Class B scheme is higher than that shown in Fig.~\ref{fig:Fig3_energy_vs_maxSF}, because each ED must expend additional energy listening for and demodulating beacon and ping messages. This additional energy expenditure, which depends on the SF used for pings and beacons, is depicted in Fig.~\ref{fig:Fig6_RxEnergy}, which shows the total energy spent by an ED per cycle, normalized by $\mathcal{P}_r$. By contrast, the WuR scheme spends negligible amount of energy in receiving, thanks to WuR properties.        

\begin{figure}
    \centering
    \includegraphics[scale=0.35]{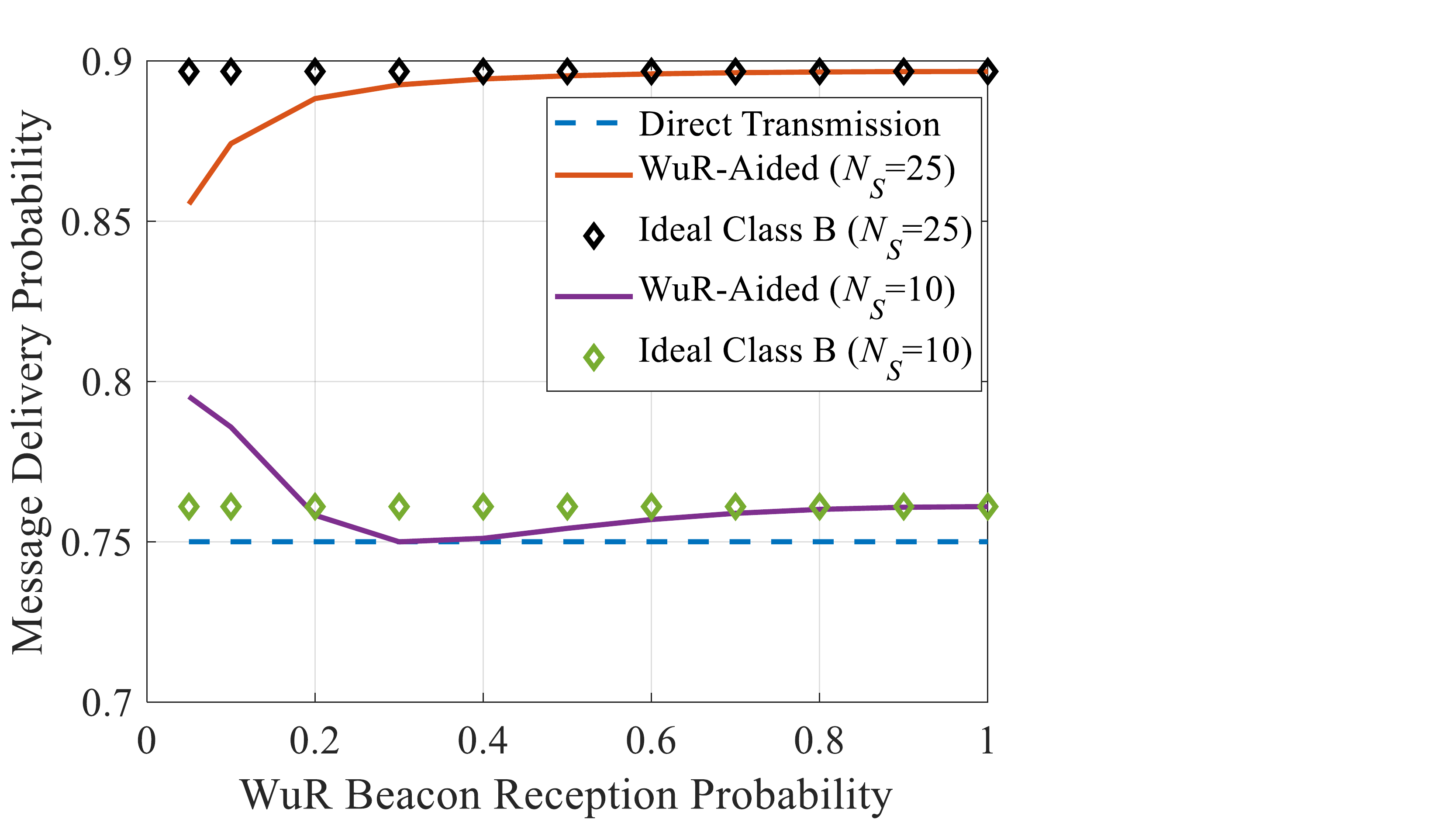}
    \caption{Delivery performance as a function of the WUB reception probability.}
    \label{fig:Fig4_MDP_vs_Pb}
\end{figure}

\begin{figure}
    \centering
    \includegraphics[scale=0.35]{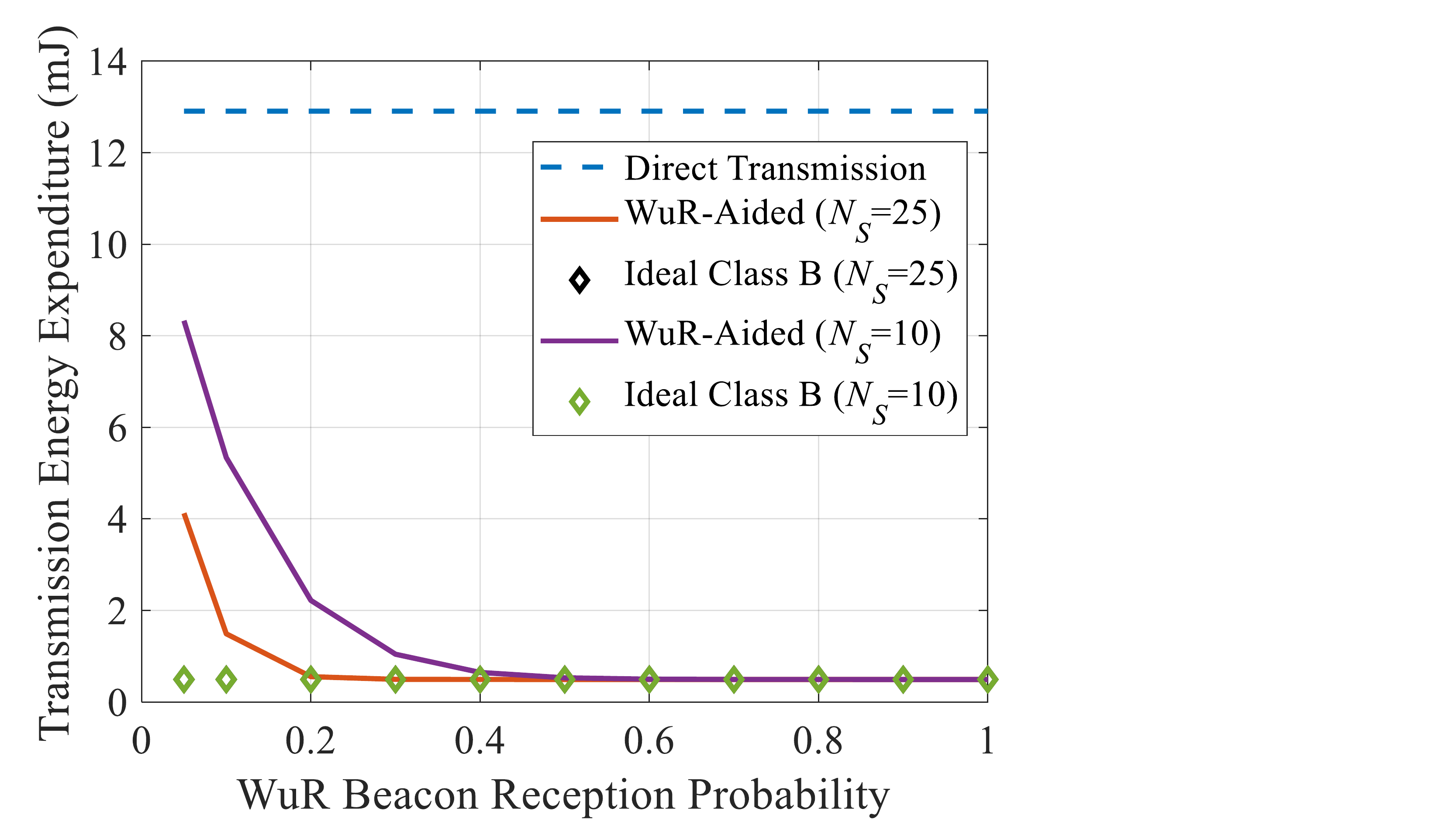}
    \caption{Transmission energy expenditure vs. WUB reception probability.}
    \label{fig:Fig5_energy_vs_Pb}
\end{figure}

In Fig.~\ref{fig:Fig4_MDP_vs_Pb}, we examine the impact of the WuR beacon reception probability on the performance of the WuR-aided scheme.  
We observe two different trends depending on the number of slots $N_s$. For $N_s \equals 25$, the performance improves with the WUB reception probability. This is a consequence of the reduced interference on the ED-to-UAV uplink. With more and more EDs waking up early due to a high beacon reception probability, they have more slots to spread their frames over, resulting in fewer collisions. Conversely, with many EDs waking up late, most frames are sent  in the last few of the $25$ slots,  encountering more collisions and subsequent loss. For $N_s \equals 10$, the curve for the WuR-aided scheme displays a non-monotonous trend. As the beacon reception probability falls below a certain point, the message delivery probability increases rather than decrease. Because $N_s$ is relatively  small, very  low beacon arrival rates lead to situations in which many EDs wake up late so as not to have enough slot to send all their messages to the UAV, sending them directly to the control station instead. Because sufficiently long random backoffs can be enforced for these transmissions, they tend to encounter fewer collisions than obtained by fitting all (or most of) the frames into the $N_s$ (or fewer slots). The associated energy expenditures are shown in Fig.~\ref{fig:Fig5_energy_vs_Pb}, which illustrates how the increased number of direct transmissions (which require higher transmit power) resulting from low beacon arrival probabilities lead to higher energy expenditure at the EDs. However, with a beacon reception rate of $40$\,\% or higher, the transmission energy requirement for the WuR-aided scheme is almost identical to the ideal Class B scheme. In practice, the requirement would actually be smaller due to the additional need of Class B to periodically receive beacons and ping messages.      

\section{Conclusion and Outlook}
Our preliminary investigation into the use of wake-up radios for UAV-aided data collection in LoRa sensor networks show that wake-up radio is an interesting alternative to the energy-inefficient Class B LoRa synchronization. In this regard, we designed and analyzed a simple random-access scheme that exploits orthogonal frequency channels and quasi-orthogonal spreading factors. Future work includes: \mbox{(A) the} incorporation of an uplink interference model that accounts for the fading on the links and the capture behavior of LoRa signals, (B) detailed modeling of the wake-up beacon reception probabilities as a function of the UAV altitude and channel characteristics, and (C) the development of a procedure to jointly allocate the SF set and number of slots with a goal of optimizing a cost metric that accounts for the message delivery probability and the energy expenditures at the sensors and the UAV.     

    


\appendices
\section{Derivation of~\eqref{eq:coll_prob}}
\label{app:coll_prob}
Consider an arbitrary ED (call it $\mathrm{E}_i$). If $\mathrm{E}_i$ wakes up in slot $W \equals i$ and has a total of $M \equals m_0$ messages to send, the probability that one of its transmissions collides with the desired frame sent in slot $s$ is equal to the probability that $\mathrm{E}_i$ also transmits a message in slot $s$. This conditional probability is $P_{\mathrm{tx}}(s|W\equals i,M \equals m_0)$, given by~\eqref{eq:cond_tx_prob}. Deconditioning on $W$ and $M$, we obtain
\begin{align} \label{eq:coll_prob_app}
P_{\mathrm{col}}(s) &= \sum_{m_0 = 1}^{M_{\max}} \sum_{i=0}^{s}P_{\mathrm{tx}}(s|W\equals i,M \equals m_0)P_M(m_0)P_W(i).
\end{align}
Observe that we must consider values of $i$ up to $s$ only, since an ED cannot send a message in slot $s$ if it does not wake up until that slot. 
Substituting the RHS of~\eqref{eq:cond_tx_prob} for $P_{\mathrm{tx}}(s|W\equals i,M \equals m_0)$ in~\eqref{eq:coll_prob_app}, and then using the fact that $P_M(m_0) = 1/M_{\max}$, we obtain
\begin{align} \nonumber
P_{\mathrm{col}}(s) 
&= \sum_{m_0 = 1}^{M_{\max}}  P_M(m_0)\sum_{i=0}^{s} \min\left\{\frac{m_0}{N(i)}, 1\right\} P_W(i) \\ 
&= \frac{1}{M_{\max}}  \sum_{m_0 = 1}^{M_{\max}} \sum_{i=0}^{s}\min\left\{\frac{m_0}{N(i)}, 1\right\} P_W(i).
\end{align}

\section{Derivation of~\eqref{eq:des_tx_prob}}
\label{app:des_tx_prob}
We are interested in finding the probability that a given message is sent in slot $s$. To that end, first condition on the events that the source of the message wakes up in slot $W \equals i$ and has a total of $M \equals m_0$ messages to send. As defined earlier, the $m_0$ messages must be spread over $N(i) \equals N_s \minus i$ slots. Now two situation can occur. If $m_0 \lteq N(i)$, then the message is guaranteed to be sent in one of the $m_0$ slots. If $m_0 \gthan N(i)$, the message is picked for transmission to the UAV with probability $N(i)/m_0$; else, it is sent directly to the control station. Combining these two cases, we can express the probability that the sender  picks the desired message for transmission to the UAV as $\min\{N(i)/m_0,1\}$. Conditioned on the event that the message is picked for transmission to the UAV, each of the $N(i)$ slots is equally likely to be selected for the message; thus, probability of choosing slot $s$ is $1/N(i)$. Thus, conditioned on  $W \equals i$ and $M \equals m_0$, the conditional probability that the message is sent in slot $s$ is  $\min\{N(i)/m_0,1\} \cdot (1/ N(i))$. Removing the conditioning on $W$ and $M$, and noting that $P_M(m_0) \equals 1/M_{\max}$, we obtain the transmission probability to be
\begin{align} \label{eq:des_tx_prob_app} \nonumber
    \mathcal{T}(s) = & \sum_{m_0=1}^{M_{\max}} \sum_{i=0}^{s} P_M(m_0)P_W(i)  \min\left\{\frac{N(i)}{m_0},1\right\} \frac{1}{N(i)}\\
    &= \sum_{m_0=1}^{M_{\max}} \sum_{i=0}^{s} \frac{P_W(i)}{M_{\max}}  \min\left\{\frac{N(i)}{m_0},1\right\} \frac{1}{N(i)}.
\end{align}

\section{Derivation of~\eqref{eq:lambda}}
\label{app:lambda}
Suppose the sender wakes up in slot $W \equals i$ and has \mbox{$M \equals m_0$} messages to send. As described in Appendix~\ref{app:des_tx_prob}, the probability that a given message is picked for transmission to the UAV is $\min\{N(i)/m_0,1\}$. Deconditioning on $W$ and $M$, and then substituting  $P_M(m_0) \equals 1/M_{\max}$ in the resulting expression,  we obtain
\begin{align} \nonumber
    \lambda &= 1 - \sum_{m_0=1}^{M_{\max}} \sum_{i=0}^{N_s-1} P_M(m_0)P_W(i)  \min\left\{\frac{N(i)}{m_0},1\right\} \\ 
    &=  1 - \sum_{m_0=1}^{M_{\max}} \sum_{i=0}^{N_s-1} \frac{P_W(i)}{M_{\max}}  \min\left\{\frac{N(i)}{m_0},1\right\}. 
\end{align}

 \balance

\end{document}